**Mimicking large spot-scanning radiation fields for proton FLASH preclinical studies with a robotic motion platform**

**Running title: Robotic motion technique for proton FLASH studies**

Fada Guan[1*], Dadi Jiang[2*], Xiaochun Wang[1*], Ming Yang[1*], Kiminori Iga[1,3], Yuting Li[1], Lawrence Bronk[1], Julianna Bronk[2], Liang Wang[2], Youming Guo[2], Narayan Sahoo[1], David R. Grosshans[2], Albert C. Koong[2], Xiaorong R. Zhu[1#], Radhe Mohan[1#]

[1] Department of Radiation Physics, The University of Texas MD Anderson Cancer Center, 1515 Holcombe Boulevard, Houston, Texas 77030, USA

[2] Department of Radiation Oncology, The University of Texas MD Anderson Cancer Center, 1515 Holcombe Boulevard, Houston, Texas 77030, USA

[3] Particle Therapy Division, Hitachi America, Ltd, 2535 Augustine Drive, Santa Clara, California 95054, USA

[*] These authors contributed equally to this work.

[#] Authors to whom correspondence should be addressed. Electronic mail: XRZhu@mdanderson.org and RMohan@mdanderson.org

**Abstract**

Previously, a synchrotron-based horizontal proton beamline (87.2 MeV) was successfully commissioned to deliver radiation doses in FLASH and conventional dose rate modes to small fields and volumes. In this study, we developed a strategy to increase the effective radiation field size using a custom robotic motion platform to automatically shift the positions of biological samples. The beam was first broadened with a thin tungsten scatterer and shaped by customized brass collimators for irradiating cell/organoid cultures in 96-well plates (a 7-mm-diameter circle) or for irradiating mice (1-cm$^2$ square). Motion patterns of the robotic platform were written in G-code, with 9-mm spot spacing used for the 96-well plates and 10.6-mm spacing for the mice. The accuracy of target positioning was verified with a self-leveling laser system. The dose delivered in the experimental conditions was validated with EBT-XD film attached to the 96-well plate or the back of the mouse. Our film-measured dose profiles matched Monte Carlo calculations well (1D gamma pass rate >95%). The FLASH dose rates were 113.7 Gy/s for cell/organoid irradiation and 191.3 Gy/s for mouse irradiation. These promising results indicate that this robotic platform can be used to effectively increase the field size for preclinical experiments with proton FLASH.

# 1. INTRODUCTION

FLASH radiotherapy refers to the delivery of radiation at ultra-high dose rates, i.e., in excess of an average dose rate of 40 Gy per second.[1] Research on and clinical use of FLASH radiotherapy are increasing nationally and worldwide,[2] largely because of the potential for FLASH radiotherapy to reduce toxicity to normal tissues while maintaining the same level of tumor control as radiation delivered at conventional dose rates (~2 Gy per minute on average).[3-8] However, the physio-chemical and biological basis for FLASH effects remains largely unknown. The mechanisms underlying the ability of FLASH radiotherapy to spare normal tissues are still being investigated but may involve inherent differences between normal tissues and tumors related to oxygenation.[9] Oxygen is critical for the cytotoxic effects of radiation, and thus well-oxygenated tissues such as the intestines are easily damaged by radiation. FLASH radiotherapy is thought to spare normoxic tissues by rapidly consuming oxygen in the targeted area, which obviates the toxic effects of oxygen in irradiated tissues. According to this hypothesis, the response of tumor tissues does not vary between conventional and FLASH dose rates because many tumors, particularly pancreatic cancer, are hypoxic.[10] Whether very high dose rates cause oxygen depletion in tissues, thereby rendering healthy tissues radioresistant, while delivering much higher biologically effective doses to tumor tissues, even in highly hypoxic areas, remains an open question. Another hypothesis of the FLASH effect is that the ultra-high dose rate radiation induces a distinct immune response.[11, 12] Because the irradiation time is much shorter in FLASH, far fewer lymphocytes would be irradiated, thereby reducing the subsequent induction of chromosomal aberrations.[11-13] However, even if fewer lymphocytes are exposed, those lymphocytes may receive a greater dose. Regardless, evidence to support the immune effects of FLASH radiotherapy is preliminary.[14, 15]

FLASH radiotherapy has the potential to herald a new era in cancer treatment if it demonstrates the same or similar treatment effects in tumors for which toxicity limits the use of curative radiation doses. Early explorations of this premise include clinical trials of electron FLASH with animals (cats and mini pigs), some of which have revealed late toxic effects.[16] A clinical trial of electron FLASH for human skin melanoma metastases was begun in 2021 (ClinicalTrials.gov identifier: NCT04986696). The FAST-01 trial (begun in 2020) has demonstrated the feasibility of proton-based FLASH for human extremity bone metastases (ClinicalTrials.gov identifier: NCT04592887).[17, 18] The newer FAST-02 trial, which was begun in 2023 and is currently open to patient enrollment, is designed to assess toxicity profiles of proton FLASH and pain relief in subjects with painful thoracic bone metastases (ClinicalTrials.gov identifier: NCT05524064).[19] Details of these and other planned FLASH clinical trials are presented elsewhere.[20-24]

Many preclinical FLASH studies have been based on modified electron therapy.[6, 25, 26] However, the short penetration range of electrons limits their use for treating deep-seated tumors. Proton therapy, as compared with conventional photon (x-rays) and electron therapy, spares normal tissues by depositing Bragg peak doses at tumor targets but lower dose at the beam entrance and no exit dose to normal tissues. If this normal tissue sparing could be further enhanced by delivering protons at ultra-high dose rates,[27, 28] FLASH proton therapy could be more potent than other conventional modalities in improving clinical outcomes.

Both particle physics-driven biological studies and clinical applications of FLASH radiotherapy require detailed characterization of the particle beams. Our team recently adapted a synchrotron-based horizontal beamline (Hitachi, Ltd., Tokyo, Japan) to deliver FLASH proton radiotherapy; details of the physics commissioning procedures and results are reported elsewhere.[29] After the commissioning process, this proton FLASH beamline was used for preclinical studies; in some of these studies, a double-scattering system was designed and mounted in the beamline for irradiation experiments based on the spread-out Bragg peak (SOBP).[30, 31] Spot-scanning irradiation has also been proposed, but this horizontal beamline is not equipped with steering magnets to control the scanning path of proton beams, and such magnets cannot be installed because of the compact design of the nozzle. Therefore, strictly speaking, the current beamline cannot be used to implement spot-scanning delivery. To resolve this issue, we report here our novel alternative method for mimicking the spot-scanning technique. In clinical spot scanning, the irradiation target is localized to a fixed position while the proton beam actively scans the irradiation target. In our alternative design, the proton beam direction is fixed along the central axis while a robotic motion platform is used to control the planar motion of the target so that the target can be scanned. Here, we describe the characterization of this FLASH proton beam involving the robotic motion technique and its use in two types of preclinical experiments. Our premise is that adapting this low-cost robotic motion platform can avoid the need for installing steering magnets while achieving sufficiently large irradiation fields to mimic the spot-scanning technique.

## 2. MATERIALS AND METHODS

### 2.1. Proton beamline

A synchrotron-based horizontal proton beamline at our institution was commissioned to deliver dose in either FLASH or non-FLASH mode and has been dedicated for preclinical studies of proton FLASH effects.[29] The increased dose rate required for FLASH delivery by our synchrotron system is accomplished by adjusting the charge in one "spill" (proportional to number of protons in a bunch per cycle) and the spill length (pulse-on time). The system can achieve several proton currents and spill lengths to deliver different dose rates and different doses per spill in FLASH mode. We chose one particular beam condition ("ID 401") for our preclinical experiments based on the stability of the beam output and precise dose control.[29]

While this beamline can deliver only a single beam energy of 87.2 MeV, various functions are made possible by several components installed in the nozzle along the beam direction. The entrance window of this nozzle is a 0.05-mm-thick stainless-steel foil ($\rho$=8.0 g/cm$^3$). A profile monitor (multi-wire ion chamber) and a reference monitor (multi-wire ion chamber) are installed next to the entrance window. The output signal from the reference monitor is fed to the amplifier of the main dose monitor for the readout of monitor units (MU) to control the beam output. An optional first scatterer, located 17 cm downstream of the entrance window, is placed next to the reference monitor. This first scatterer is a 0.3-mm-thick tungsten film ($\rho$=19.3 g/cm$^3$) that can be used to broaden the beam profile laterally. The next component is a large-hole collimator made of brass ($\rho$=8.07 g/cm$^3$); the 10-cm inner diameter of this collimator allows the beam to pass through to minimize interactions with its metal parts. This large-hole collimator is installed vertically; it is also used as the stand for alignment of experimental devices, and the center of the hole at the exit surface is defined as the isocenter for experimental setups. The distance between the experimental isocenter and the entrance window is approximately 55 cm. The isocenter plane was chosen as the reference plane for characterizing the dose and dose rate of this beamline. The dose rate depends strongly on the use of beam shaping devices and the choice of experimental setup; as such, in this report we sought to minimize confusion by specifying the experimental condition and location at which the dose rate is reported. This FLASH beamline was modeled with the Geant4 Monte Carlo simulation toolkit[32, 33] with Gafchromic™ film (Ashland Specialty Ingredients, G.P., Wilmington, DE) and Advanced Markus chamber (PTW, Freiburg, Germany) measurements performed for validation (**Figure 1**).

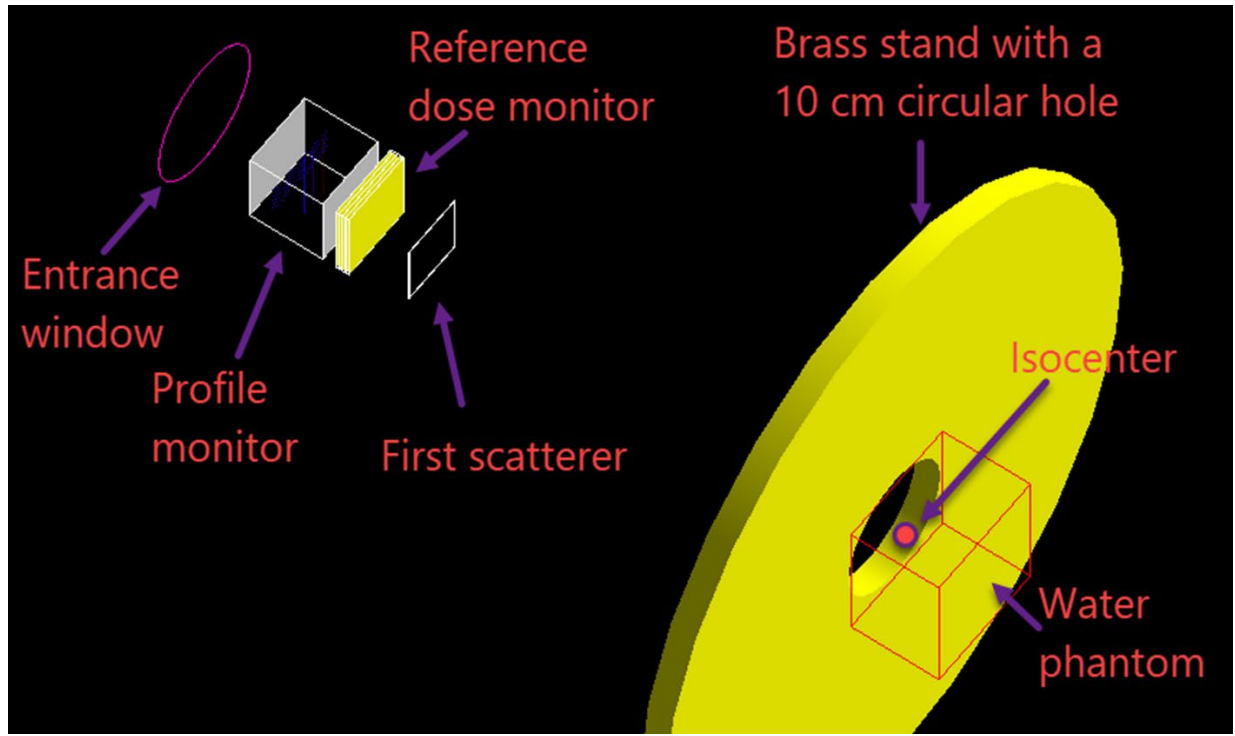

**Figure 1.** Schematic of Monte Carlo modeling of the proton FLASH beamline.

**2.2. Time structures in FLASH and non-FLASH modes**

Understanding the time structure of the beam delivery is crucial for reporting the instantaneous dose rate and average dose rate of this modified system. Protons from a synchrotron are delivered in a quasi-continuous mode. The operation cycle of our synchrotron is 2 s regardless of the dose rate. Simplified time structures for non-FLASH and FLASH mode in our synchrotron are compared in **Figure 2.** In non-FLASH mode, the pulse-on length is 0.5 s to deliver protons in a full spill, and the idle time before the start of the next pulse is 1.5 s. The duty cycle (ratio of pulse-on time and operation cycle) in the non-FLASH mode is 25%. In contrast, in FLASH mode, the pulse-on length is shortened to approximately 0.1 s at maximum (ID 401 in this study) to deliver a full spill with a duty cycle of only 5%. In FLASH mode, the duty cycle can be lowered further when only a partial spill is delivered. The pulse structures of a full spill and a partial spill recorded by an oscilloscope (Tektronix, TDS 3014B) are compared in **Figure 3**. The fluctuation in these pulse structures indicates that the instantaneous proton fluence rate and dose rate vary over time within a pulse. In general, pulse height decreases with time. Notably, the MU and pulse-on time of a full spill vary in different deliveries. Therefore, we recorded the MU and spill length (in ms) for each FLASH dose delivery in our experiments.

In FLASH mode, a typical full spill of protons corresponds to approximately 1450 MU with a pulse-on time of approximately 0.1 s. In the partial spill shown in **Figure 3B**, the pulse-on time was 41.8 ms for 807.4 MU. The pulse-on time vs. output MU in a single spill (full or partial) from a set of data collected on the same day and the second order polynomial fit of the data (adjusted $R^2$ = 0.965) are illustrated in **Figure 4**. The superlinear increase of recorded pulse-on time with output MU indicates that the average MU delivery rate (i.e., MU/ms) in a pulse decreases when the machine MU is set to a higher value. Because the delivered dose is proportional to its MU, the average dose rate within a pulse also decreases with MU. The large signal fluctuation within a pulse (e.g., **Figure 3**) makes acquiring the instantaneous dose rate challenging. In this report, to minimize confusion, the dose rate is expressed as the "average" value within a pulse, i.e., delivered dose divided by the pulse-on time length. Because the main dose monitor is an open-air ion chamber, the output MU needs to be corrected to be the value at the standard environmental conditions of temperature at 22°C and pressure at 101.33 kPa (~1 atmosphere). All the output MU values in this report refer to the standard values with these temperature and pressure corrections.

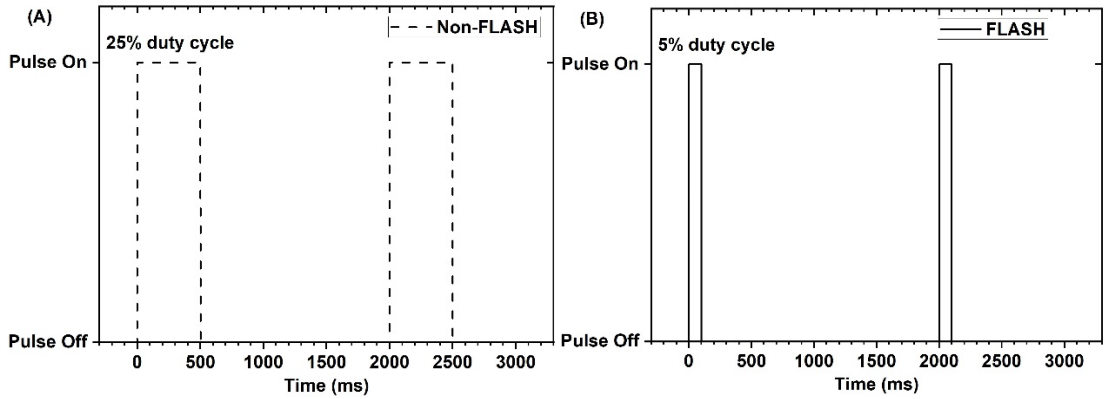

**Figure 2.** Simplified time structures of the dose delivered by the synchrotron used in (A) conventional dose rate (i.e., non-FLASH) and (B) FLASH modes.

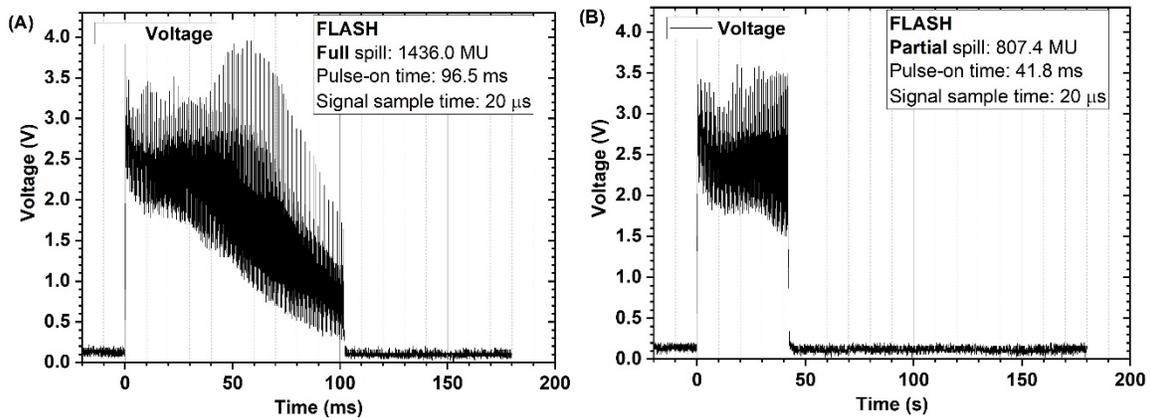

**Figure 3.** Examples of FLASH pulse structures for (A) a full spill and (B) a partial spill.

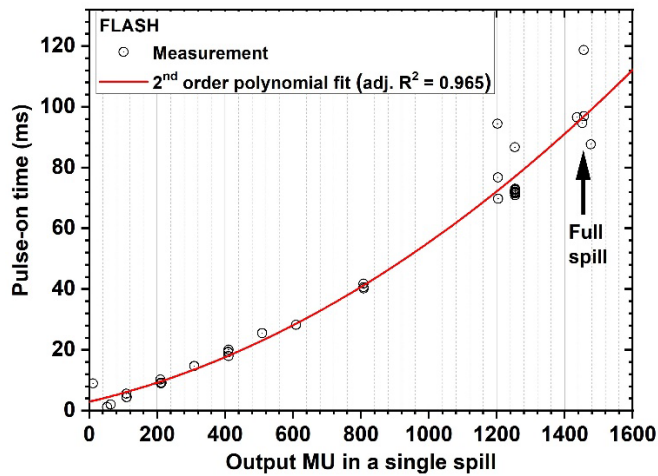

**Figure 4.** Pulse-on time as a function of the output MU in a single spill (partial or full) in FLASH mode. The second order polynomial fit of the experimental dataset shows a superlinear increase of recorded pulse-on time with output MU.

The output MU and pulse-on time are fundamental quantities for calculating the dose and dose rate, and so in this report they are provided together with the average dose rate for each setup. In this study, in FLASH mode, most of the desired dose (<20 Gy) can be delivered within a single pulse (full or partial spill), and thus the average dose rate can reasonably be expressed within a pulse. In non-FLASH mode, however, the MU delivery rate is much lower. Because strong ion recombination takes place in the main dose monitor (open-air ion chamber) in FLASH mode, with the same number of protons delivered, the output MU (proportional to collected charge in the dose monitor) ratio was found to be 0.928 between the FLASH and non-FLASH mode. With the same setup, the ratio of average dose rate "during duty cycle" between the FLASH and non-FLASH mode was approximately 131.8. If the average dose rate of non-FLASH mode is calculated considering the idle time (which is 3 times the pulse-on time), the above ratio could be 527.2 (=131.8 × 4). In this report, we present only the average dose rate of FLASH mode, with its non-FLASH counterpart acquired by scaling accordingly. For example, if the FLASH dose rate averaged within a pulse is 100 Gy/s, its non-FLASH dose rates averaged within a pulse and within a full cycle would be 0.76 and 0.19 Gy/s, which are still higher than the conventional dose rate of ~2 Gy/min (0.03 Gy/s) used in current clinical applications.

## 2.3. Measuring the original and broadened beam spots

The original beam spot is narrow and thus is not suitable for irradiating large targets. The beam profile can be broadened laterally by inserting an optional first scatterer into the beamline. The setups for both the narrow and broadened beams, measured with Gafchromic™ EBT-XD films in the isocenter plane, are shown in **Figure 5**. EBT-XD films were chosen for their extended effective range of dose response (0.4 to 40 Gy) and their dose-rate-independence within the range of dose rates in our proton FLASH system.[34]

The depth dose along the central axis from the broadened beam spot was measured with EBT-XD film and with an Advanced Markus chamber at a few locations in a plastic water phantom.[29] The measurement data were used to validate and calibrate the Geant4 Monte Carlo simulations. For the original narrow beam, only film measurements and Monte Carlo simulation were used to acquire depth dose data because the central area of the original beam is smaller than the diameter (5.0 mm) of the sensitive volume of the Advanced Markus chamber. Ion chamber measurements were corrected for ion recombination for ultra-high dose rates.[35]

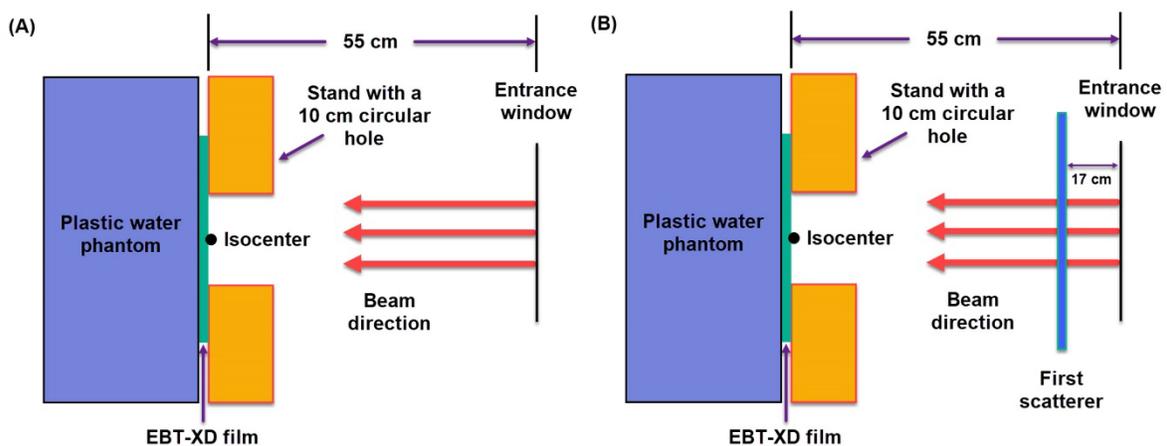

**Figure 5.** Schematic of setups for film measurement of spots from (A) the original narrow beam and (B) the broadened beam after passing through the first scatterer.

## 2.4. Robotic motion technique for increasing the radiation field size

The field size from a single beam, even when broadened by the scatterer, is not large enough to cover biological samples in preclinical experiments. Two possible solutions can be used to extend the field size: (1) using a double-scattering system to further broaden the beam size, as implemented by Titt *et al.* for this nozzle,[30] and (2) using a scanning technique to deliver multiple beamlets to different locations to form a large field. In clinical applications, steering magnets are used to control the beam direction in the scanning mode. Because installing steering magnets in this horizontal beam nozzle was not feasible, we applied a robotic motion technique to move the target location between beam deliveries to mimic the spot-scanning technique and extend the field size to cover larger targets. The workflow for this process is illustrated in **Figure 6**. The narrow beamlet is broadened by the first scatterer and then collimated by a small-hole collimator (made of brass, 20 cm × 20 cm × 2 cm). A collimated circular beam is used for in vitro experiments with cells or organoids cultured in 96-well plates; a collimated square beam is used for in vivo experiments involving irradiation of small animals.

To meet these goals, a dedicated experimental platform was designed that includes a self-leveling laser positioning system and a robotic motion platform (Rotrics, Shenzhen, China). A step-and-shoot beam delivery strategy was used to irradiate the biological samples. The motion pattern was written in the programming language G-code, and the code was sent from a computer to the robotic platform to control its motion. A webcam was also installed in the nozzle to monitor the motion of the target along the predefined motion pattern for accurate dose delivery to the appropriate location.

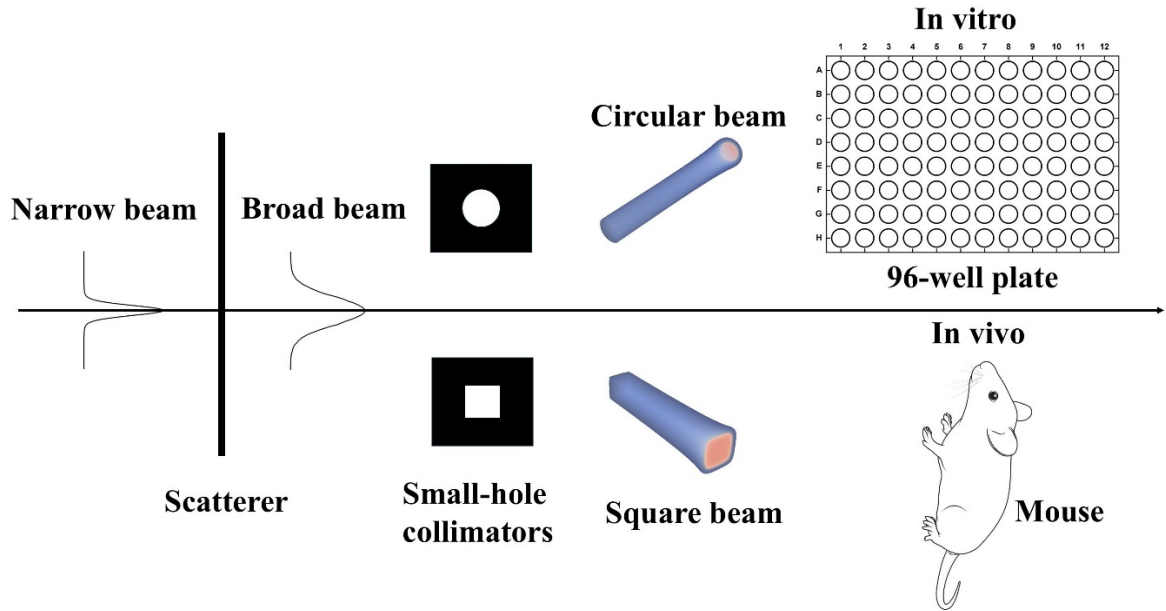

**Figure 6.** Schematic illustrating the workflow for experiments involving increased irradiation field size. Mouse image from Tyler E, Kravitz L. Scidraw. https://doi.org/10.5281/zenodo.3925901 (2020).

*2.4.1. Setup for in vitro high-throughput experiments*

The experimental setup involving a high-throughput 96-well plate mounted on the robotic motion platform with laser localization is shown in **Figure 7**. A 7-mm inner diameter collimator was used to match the inner diameter of a well. This experimental platform has been used to irradiate both cells and organoids.[36] After the dose is delivered to a specified well, the 96-well plate is moved to the next target well. This step-and-shoot procedure is repeated until all target wells are irradiated. To avoid leakage of medium, the vertically oriented 96-well plate was sealed during cell/organoid irradiation using a foil (AlumaSeal II Adhesive Sealing Foil for Plates, Research Products International).

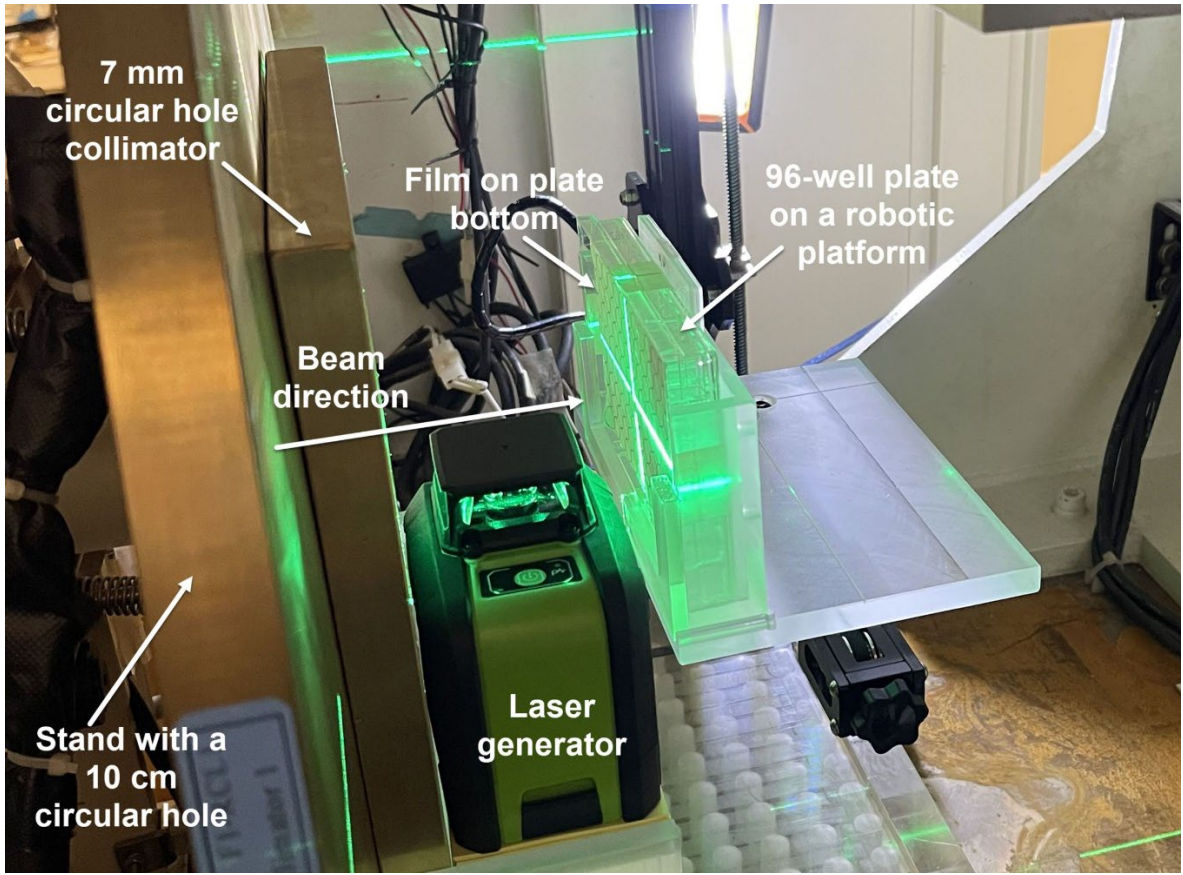

**Figure 7.** Image of the robotic motion platform and a self-leveling laser positioning system for high-throughput irradiation of cells or organoids in culture.

The spatial and dosimetric accuracy of dose delivered to the wells was validated with a piece of EBT-XD film aligned with and attached to the bottom of the 96-well plate (**Figure 8**). Film-measured dose profiles were compared with Monte Carlo-calculated profiles. A 7-cm air gap was present between the end of the 7-mm circular hole collimator and the bottom of the 96-well plate, but no extra buildup was used before the plate. The plates are made of polystyrene ($\rho$=1.09 g/cm$^3$), and the wells are ~1.2 to 1.3 mm thick at the bottom. Irradiations were done with the entrance dose so as to exclude the effect of linear energy transfer (LET) on biological responses as in the Bragg peak region from FLASH dose rate irradiation. Dose rate was the physical quantity of interest in the current experimental design; if both dose rate and LET effects are to be investigated, a multi-step range shifting device or compensator can be used as we described elsewhere.[34, 37-39]

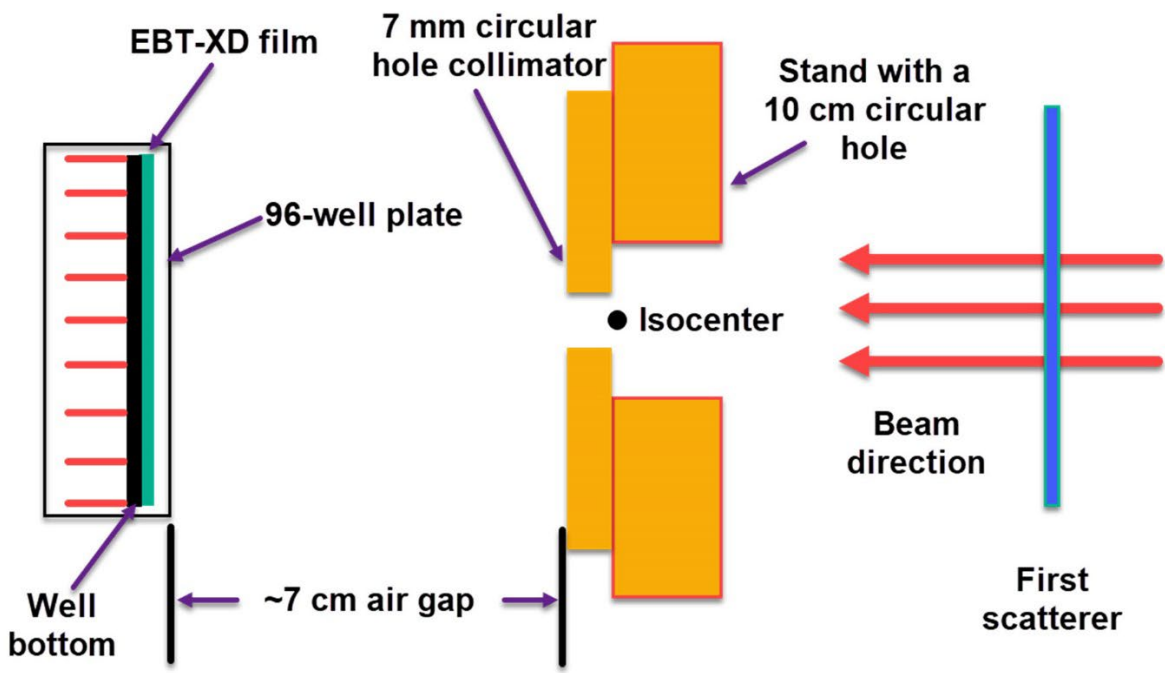

**Figure 8.** Schematic illustration of validation film attached to the bottom of a 96-well plate with a 7-cm air gap between the end of the 7-mm circular hole collimator and the bottom of the plate.

*2.4.2. Setup for in vivo small-animal irradiations*

For mouse irradiation experiments, a collimator with a 1 cm × 1 cm square hole is used to match a typical tumor size. For irradiation of larger targets, such as the entire abdomen of a mouse, several square fields can be patched to form a large field, a procedure that can also be achieved using the robotic motion platform. The setup for irradiating a mouse abdomen, with a custom holder for the mouse and an anesthesia device, is shown in **Figures 9** and **10**. The distance between the end of the square-hole collimator and the mouse holder is 2.7 cm, and a ~1-cm air gap is present between the end of the collimator and the abdomen of the mouse. The mouse abdomen is ~1.3 to 1.5 cm thick, and the entire abdomen area is ~3 cm × 3 cm. The spatial and dosimetric accuracy of dose delivery was validated using a piece of EBT-XD film attached to the back of the mouse.

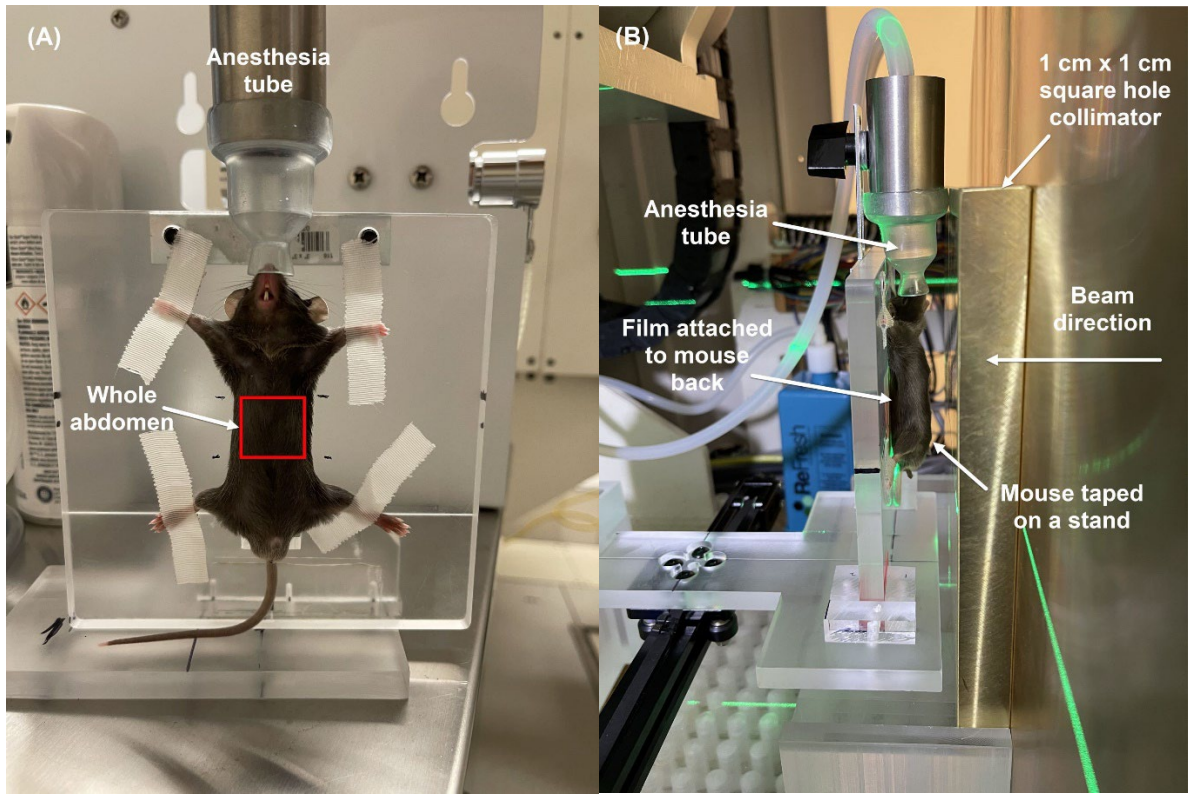

**Figure 9.** (A) Front view and (B) lateral view of the mouse setup with the robotic platform and a custom holder for the mouse and the anesthesia device. A ~1-cm air gap is present between the end of the collimator and the abdomen of the mouse. The mouse abdomen is ~1.3 to 1.5 cm thick; the whole abdomen area is ~3 cm × 3 cm.

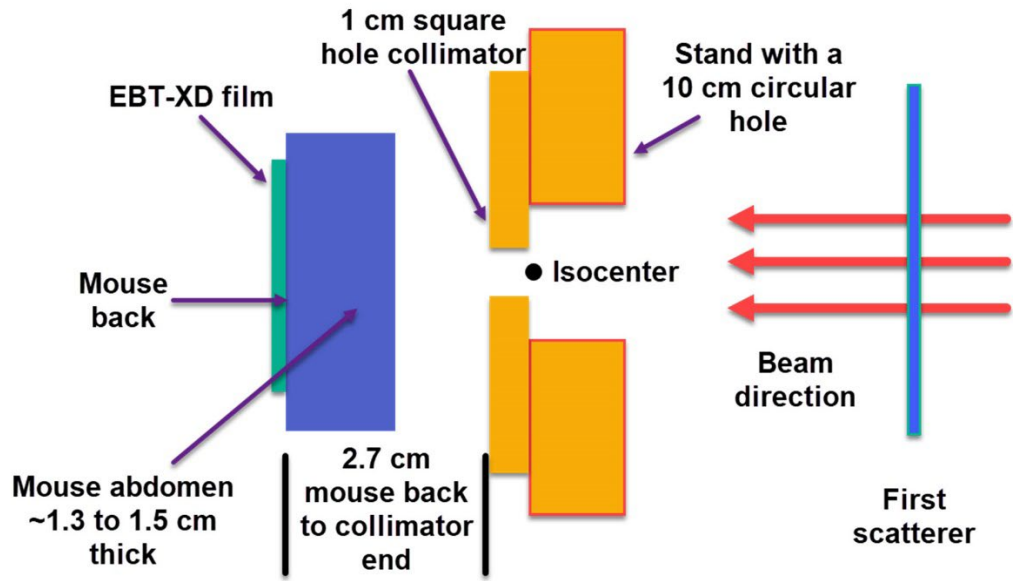

**Figure 10.** Schematic illustrating the validation film attached to the back of a mouse.

## 2.5. Film scanning and dose analysis

The EBT-XD films were scanned with an EPSON Expression 10000XL flat-bed scanner (Epson America, Inc., Los Alamitos, CA) at 24 to 48 hours after irradiation at a resolution of 254 dpi (0.1 mm × 0.1 mm per pixel) in the RGB mode with 16-bit color depth per channel. The film images were saved in the tagged image file format (.TIFF). Film images were analyzed with ImageJ software[40] (version 1.53c). EBT-XD films have been calibrated in both proton FLASH and non-FLASH modes and found to be dose-rate-independent within the range of dose rates in our proton beamline.[34] The dose map of each film was acquired from its corresponding dose calibration curve (i.e., converting optical density to dose).

## 2.6. Comparing film measurements and Monte Carlo calculations

In all cases, the films used were marked on four sides to localize the center of the beam. Each film in each experimental setup was aligned by using the laser system to ensure that the film center coincided with the beam center. Each experimental setup was also modeled by Geant4 Monte Carlo simulation. The film-measured dose distribution was compared with the Monte Carlo-calculated results, and 1D gamma analysis was performed for each setup using the open-source software *PyMedPhys* (version 0.39.3)[41] (https://pypi.org/project/pymedphys/). The film measurement results were used as the reference, and Monte Carlo calculations as the evaluation. Because dose delivery in this study was from a small field, we set more stringent criteria for gamma analysis, specifically 2%/1 mm. The dose threshold value was set to 2% of the maximum dose in the reference dataset for the global gamma index ($\gamma$) calculation to evaluate the accuracy of low dose tails as well. In addition, field width, flatness, and symmetry of small-field dose profiles were calculated using an in-house Python script as defined by Varian Medical Systems (Palo Alto, CA).

## 3. RESULTS

### 3.1. Original narrow and broadened beams

The film images measured in the isocenter plane for the original narrow beam and the broadened beam after passing through the first scatterer (**Figure 5**) are shown in **Figures 11A and 11B**. In the film measurement of the original narrow beam, a partial spill of 209.9 MU was delivered with a pulse-on time of 9.8 ms; in the film of the broadened beam, a partial spill of 1204.1 MU was delivered with a pulse-on time of 68.3 ms.

The crossline (horizontal) dose profiles (normalized by the output MU, in Gy/MU) from film measurements and Monte Carlo calculations are compared in **Figure 11C**. The gamma pass rates were 100% for the narrow beam and 99.8% for the broadened beam. The full width at half maximum (FWHM) of these two beam profiles from Monte Carlo calculations was 8.0 mm for the narrow beam and 22.0 mm for the broadened beam. The average dose rates (dose/pulse-on time) at the beam center were 1670 Gy/s for the narrow beam and 190 Gy/s for the broadened beam (**Figure 11D).** Although the first scatterer greatly reduces the dose rate (to 11.4%), the dose rate of the broadened beam was still higher than the common FLASH threshold of 40 Gy/s.

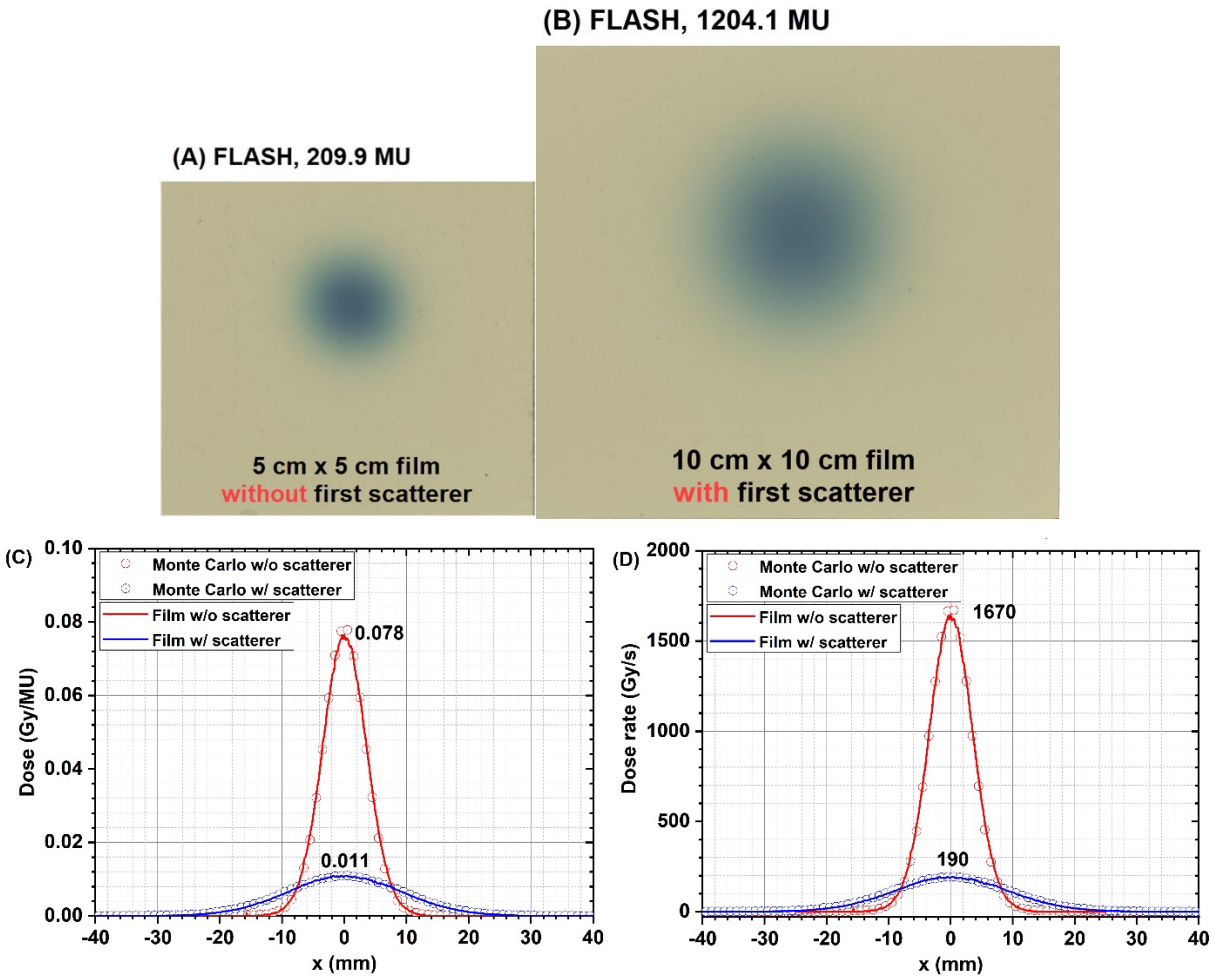

**Figure 11.** Comparison of the original narrow beam and the broadened beam with the use of the first scatterer. The film measurement was obtained in the isocenter plane.

We also evaluated the average dose rate from a full spill (1450 MU, pulse-on time 100 ms) of Bragg curves along the central axis, with and without the first scatterer (**Figure 12)**. Dose rate was calculated from the Monte Carlo-generated dose and a pulse-on time 100 ms of a full spill. The entrance dose rate for the original narrow beam was ~1000 Gy/s with a full spill of 1450 MU, which is lower than the dose rate of 1670 Gy/s in **Figure 11D** obtained from a partial spill beam of 209.9 MU. Because the average dose rate of a spill decreases with MU (**Figure 4**), the dose rate from a full spill will be lower than that of a partial spill. The Bragg peak dose rate of the original narrow beam was as high as 4500 Gy/s. In contrast, the Bragg peak dose rate of the broadened beam was reduced to 660 Gy/s. Use of the first scatterer also reduced the range of the proton beam (**Figure 12**).

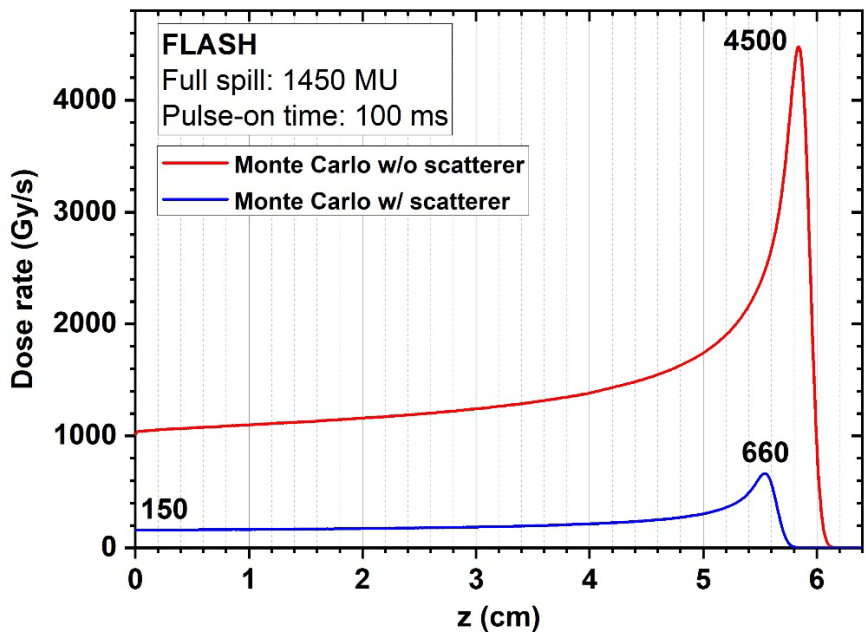

**Figure 12.** Average dose rate Bragg curves along the central axis from the original narrow beam and the broadened beam with a full spill (1450 MU, pulse-on time 100 ms). Dose rate was calculated from the Monte Carlo–generated dose and the pulse-on time 100 ms of a full spill.

## 3.2. Collimated circular beam for in vitro experiments

Neither the original narrow beam nor the broadened beam was suitable for irradiation experiments that require conformal dose distribution to a target geometry. Thus to irradiate a single well in the 96-well cell culture plate, a 7-mm circular hole collimator was used to shape the broadened beam profile to conform to the inner diameter of that well, and a film was attached to the bottom of the 96-well plate (with a distance between the end of the circular collimator and the film of ~7 cm [**Figure 8**]). The film image of a collimated circular beam is shown in **Figure 13A,** and the crossline dose profiles from the film measurement and the Monte Carlo simulation are compared in **Figure 13B**. The film was irradiated by 1612.1 MU with a pulse-on time of 104 ms, made of a full spill of 1450 MU with 96.5 ms and a partial spill of 162.1 MU with 7.5 ms. The derived average dose rate profiles from the film measurement and the Monte Carlo simulation are shown in **Figure 13C**. The average dose rate at the central axis (CAX) point according to the Monte Carlo simulation was 113.7 Gy/s. The gamma pass rate between the film measurement and the Monte Carlo simulation was 99.3%. Other radiation field profile–related parameters were also calculated for the Monte Carlo–generated profile. The field width was 8.0 mm (distance between 50% to 50% of CAX dose). The penumbra (distance between 80% and 20% of CAX dose) was 0.8 mm on either side. The symmetry was 0.6% between the left and right half of the dose profile. From –3 to 3 mm, the dose ranged from 94.2% to 100.6% of the CAX dose, and the flatness was 3.3%. From –2 to 2 mm, the dose ranged from 98.7% to 100.6% of the CAX dose, and the flatness was 1.0%. These results indicate that the cells or organoids in a well could receive a uniform dose.

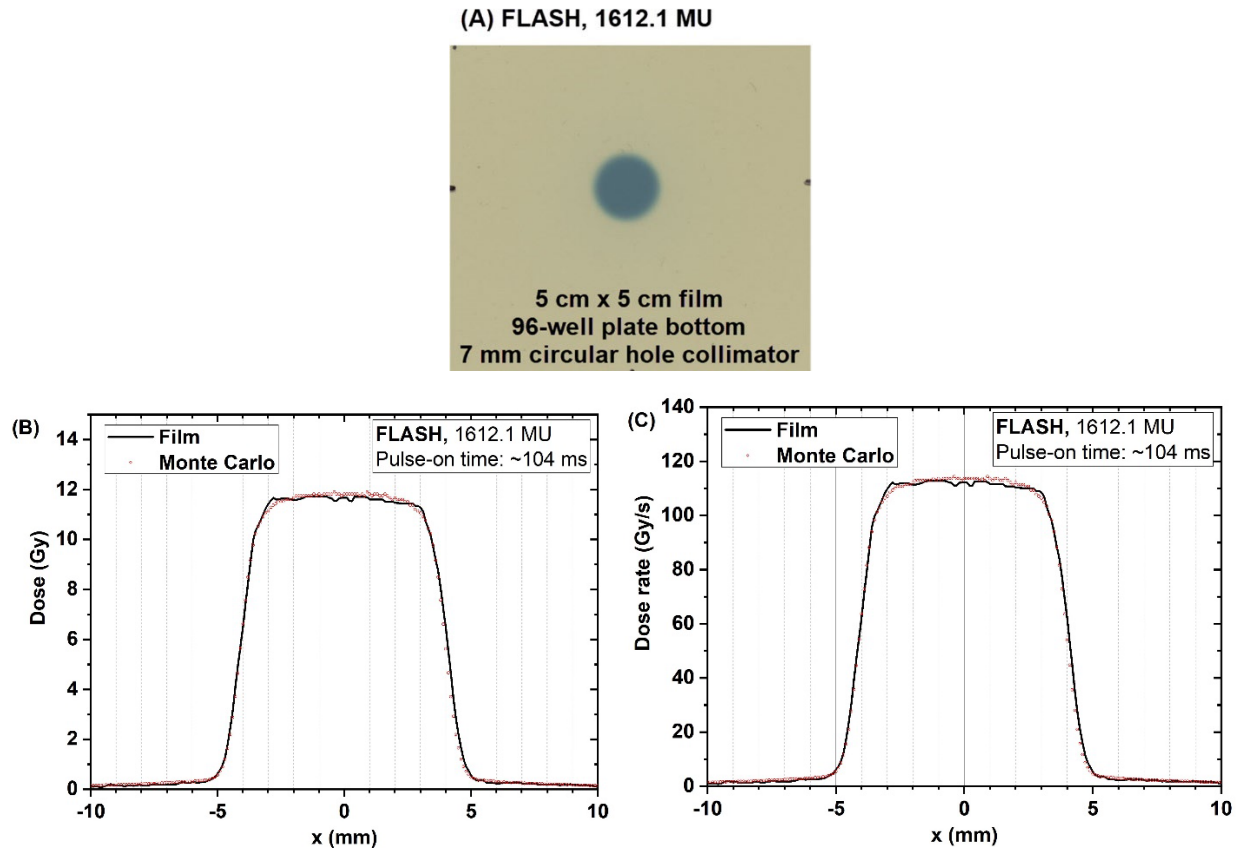

**Figure 13.** (A) Film image of a collimated circular beam irradiated by 1612.1 MU proton FLASH. (B) Crossline dose profiles and (C) average dose rate profiles from the film measurement and Monte Carlo simulation.

### 3.3. Collimated square beam for in vivo experiments

A 1-cm square hole collimator was used to shape the broadened beam profile to conform to a typical tumor size in a mouse. A film was attached to the back of a mouse bearing a tumor, and the distance between the end of the square collimator and the film was 2.7 cm (**Figure 10**). The film image of the collimated square beam is shown in **Figure 14A**. The crossline dose profiles from film measurement and Monte Carlo simulation are compared in **Figure 14B**. The film was irradiated by a partial spill of 1253.6 MU with a pulse-on time of 72.5 ms. The derived average dose rate profiles from film measurement and Monte Carlo simulation (**Figure 14C**) indicate that the average dose rate at the central axis point was 191.3 Gy/s in the Monte Carlo simulation. The gamma pass rate between the film measurement and the Monte Carlo simulation was 95.8%. Other radiation field profile–related parameters calculated for the Monte Carlo–generated profile were as follows. The field width was 10.6 mm; the penumbra was 0.5 mm on either side; and the symmetry was 0.9% between the left and right half of the dose profile. From –4 mm to 4 mm, the dose ranged from 96.2% to 100% of the CAX dose, and the flatness was 2.4%. These results indicate that a typical tumor in a mouse could receive a uniform dose laterally. As shown in **Figure 12**, the depth dose increased slowly in the first 2 cm, which is deep enough to maintain a nearly constant depth dose in the mouse abdomen (~1.3 to 1.5 cm thick).

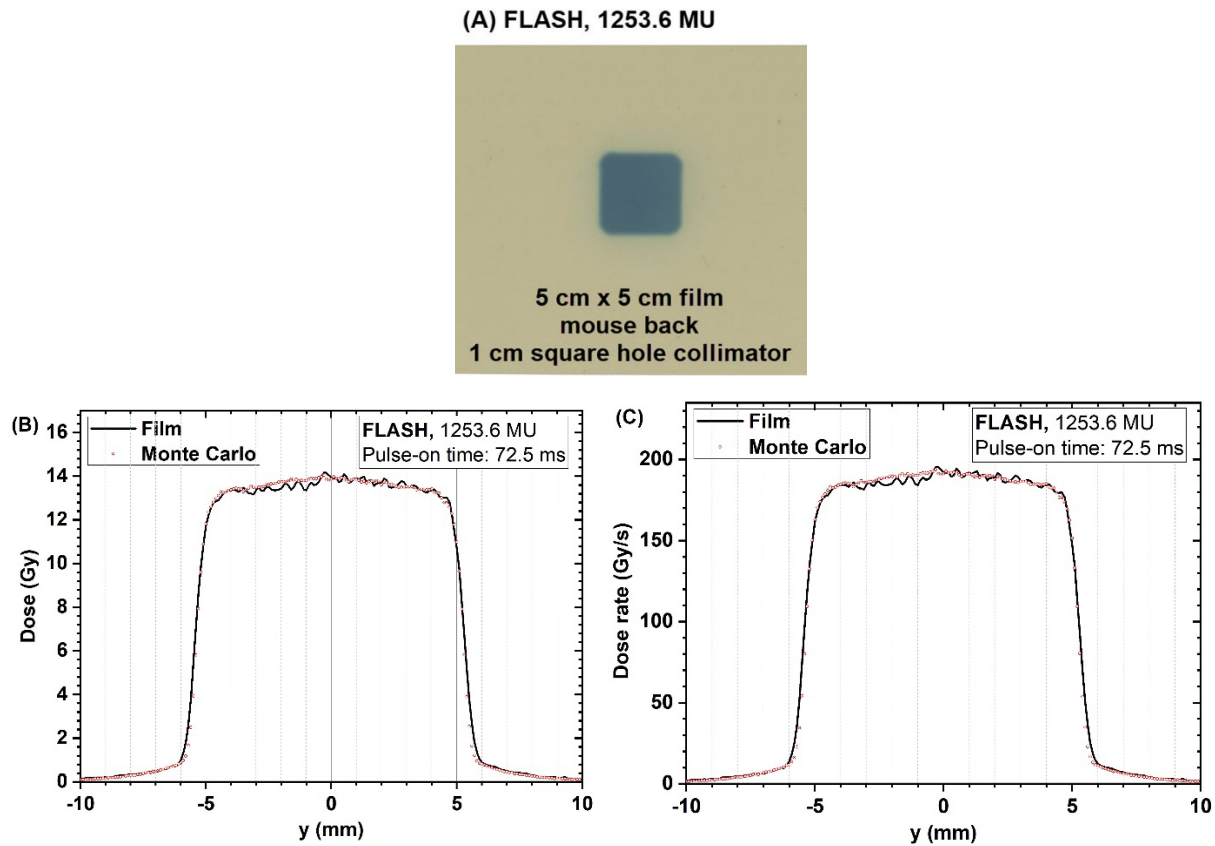

**Figure 14.** (A) Film image of a collimated square beam irradiated by 1253.6 MU proton FLASH. (B) Crossline dose profiles and (C) average dose rate profiles from film measurement and Monte Carlo simulation.

### 3.4. Multiple circular beam fields for in vitro experiments

An image of a film with multiple circular beam spots (separated by 9 mm) attached to the bottom of the 96-well plate on the robotic platform is shown in **Figure 15A**. The spots in the top half of the film were from FLASH irradiation and those in the bottom half were from non-FLASH irradiation. The crossline (horizontal) dose profiles across row C of the 96-well plate from film measurement and Monte Carlo simulation are shown in **Figure 15B**. The gamma pass rate was 98.4%, indicating good matching between film measurement and Monte Carlo calculation and validating the high accuracy of spatial positioning and dose delivery.

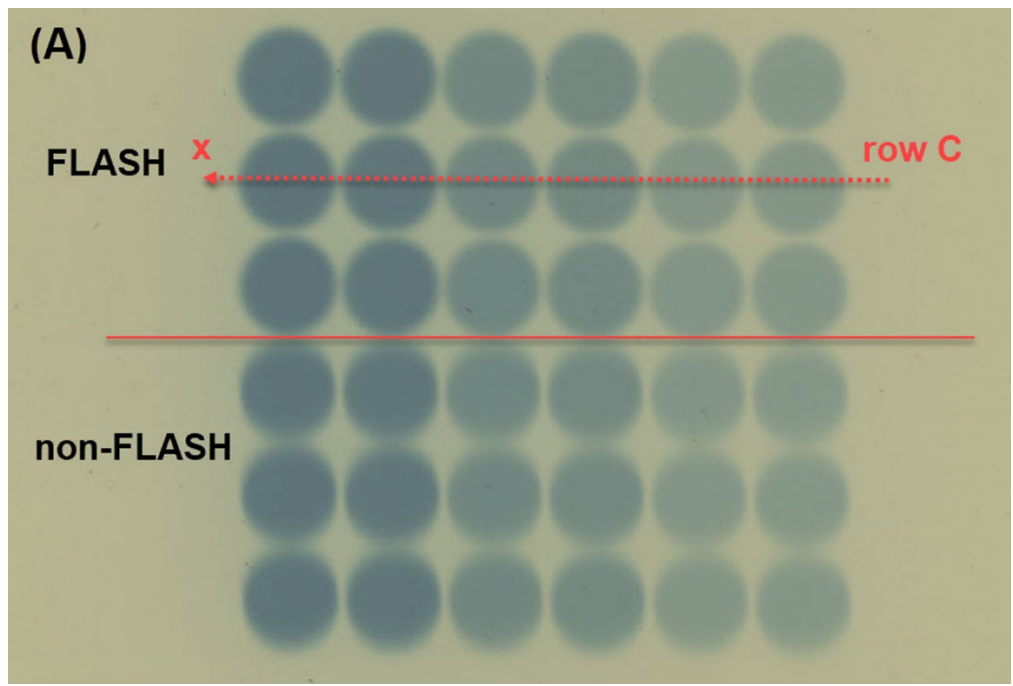

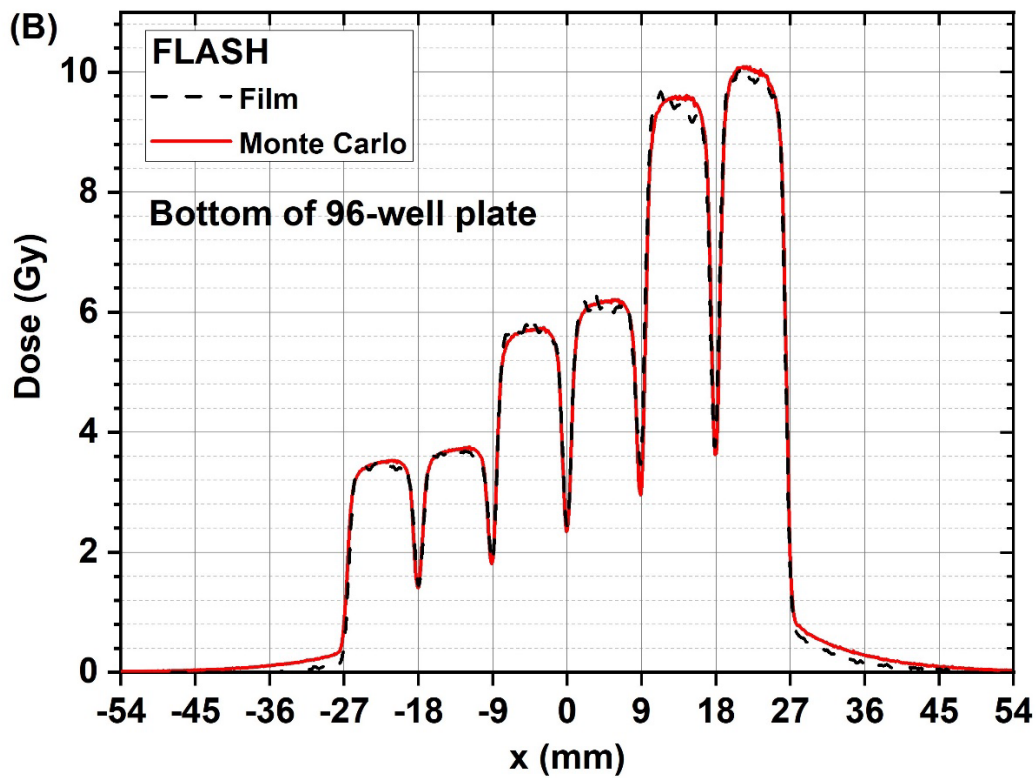

**Figure 15.** (A) Film image of multiple circular beam spots attached to the bottom of a 96-well plate with 9-mm spot spacing in the motion pattern. (B) Crossline dose profiles across row C from film measurement and Monte Carlo simulation.

3.5. Patched square beam fields for in vivo experiments

An image of a film with patched square beam fields on the back of a mouse undergoing abdominal irradiation is shown in **Figure 16A**. Four 1 cm × 1 cm beam fields (with 10.6-mm spacing vertically and horizontally) were delivered on the robotic platform. The inline (vertical) dose profiles across the center of the two patched square fields on the left side of the mouse from film measurement and Monte Carlo simulation are shown in **Figure 16B**. The gamma pass rate was 94.1%. Parameters from the Monte Carlo–generated profile were as follows. The field width was 21.2 mm; the penumbra was 0.5 mm on either side; the flatness of the profile was 2.1%; and the symmetry was 0.6%. Notably, the small diamond-shaped cold spot present at the center of the four patched fields is one of the limitations of the current experimental design. This is caused by the chamfer at each corner of the square-hole collimator.

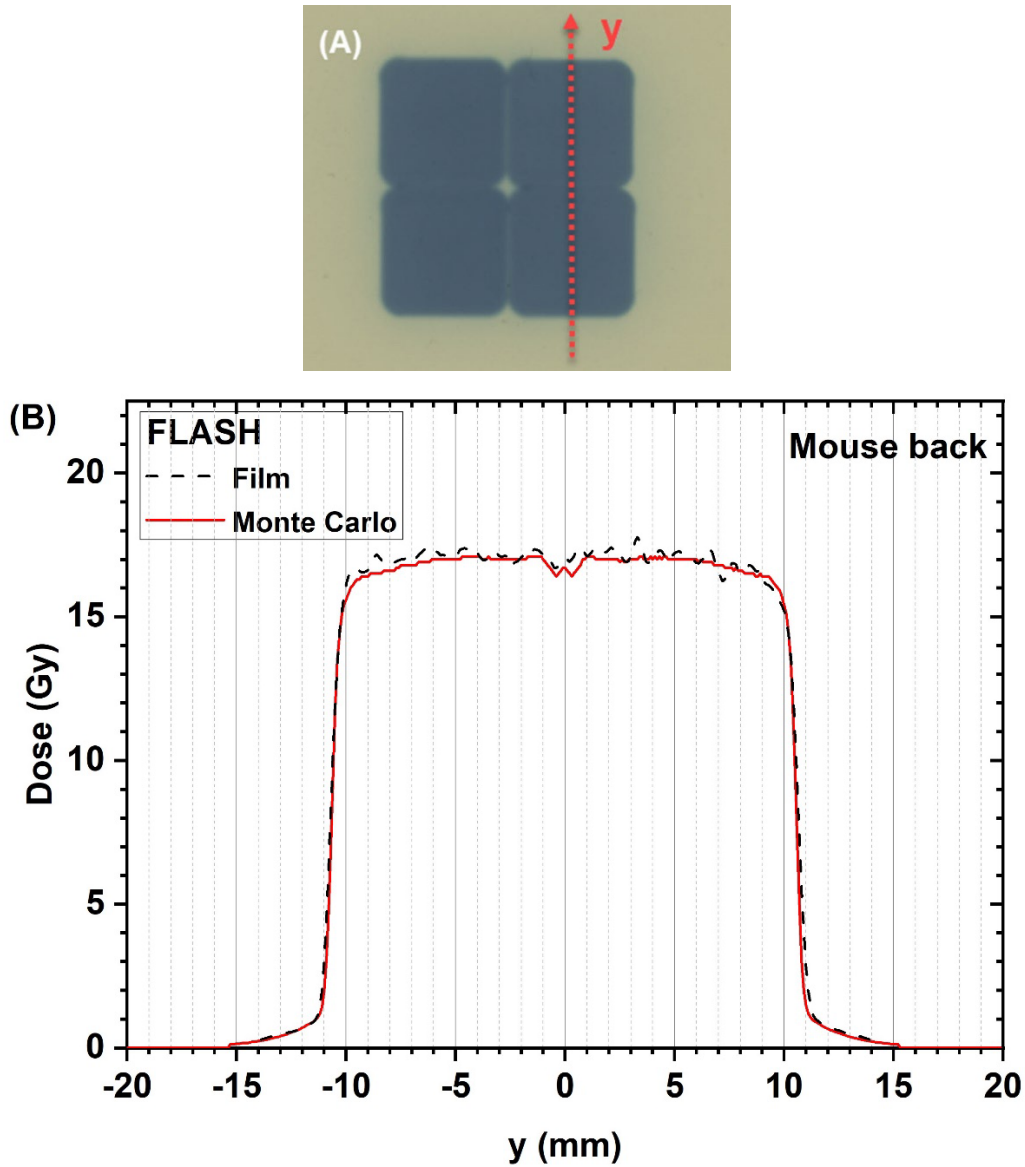

**Figure 16.** (A) Film image from the back of a mouse undergoing abdomen irradiation. 10.6-mm spacing was used in the motion pattern of the robotic platform between 1 cm × 1 cm square fields. (B) Inline (vertical) dose profiles across the center of two square-field patches on the left side of the mouse.

## 4. DISCUSSION

Given that scanning beams are required for the current state-of-the-art technique of intensity-modulated proton therapy (IMPT), many newly built proton therapy centers are solely equipped with scanning nozzles. The former passive scattering technique is gradually being eliminated in proton therapy. If ultra-high dose rate proton FLASH is applied in clinics, scanning beams would be the first choice. However, before proton FLASH is used in clinical applications, understanding the biological mechanisms underlying scanning proton FLASH is paramount. Because the experimental FLASH beamline at our proton center is not equipped with a set of scanning magnets, we developed an alternative strategy in which a robotic motion platform is used to control the motion of the target to achieve the extended radiation field size for mimicking the spot-scanning technique.

When steering magnets are not available to implement beam scanning, using a robotic motion platform with a laser localization system has several advantages. The first is the considerably lower cost of such a platform relative to installing a complicated steering magnet system. Second, we have verified that this robotic system is highly accurate for target positioning to ensure the spatial accuracy of beam delivery. Third, use of the robotic platform can improve experimental efficiency by eliminating the need to manually align the target between different beam deliveries. As an example, our use of the robotic motion platform facilitated motion control of 96-well plates so that a specified well center could be accurately aligned to the beam center. Fourth, both spatial and dosimetric accuracy could be validated by attaching film to the target in the experimental setup (either the plate or the mouse). Fifth, applying a step-and-shoot irradiation pattern could result in rapid irradiation of several cell culture wells or a relatively large area on a mouse. We conclude from these features that use of this robotic platform can achieve equivalent target dose coverage by mimicking the active scanning technique.

The current proton FLASH experimental design is also subject to some limitations. First, only one beam energy (87.2 MeV) is available for this proton FLASH beamline, and thus the short penetration range means that only experiments involving cell cultures or small animals are possible at this time. Second, film images showed that the dose profile in each well was not uniform (**Figure 15**) because of the superposition of dose tails from adjacent beam spots. One possible solution for this shortcoming could be to increase the spacing between the irradiated wells, e.g., irradiating every other well. Third, the efficiency of the experiments was much lower than the use of a real spot-scanning technique. Even though the motion platform can move quickly (within seconds) after the dose is delivered to one location (e.g., a well), the proton FLASH system needs at least 30 to 60 seconds to prepare for the next beam delivery. Thus compared with a true scanning technique for irradiating a large area, the actual time needed for our current experiments was much longer.

Nevertheless, our current findings can be extended to other future studies. As an example, the current study focused only on dose rate and its biological consequences; we used the entrance dose along the Bragg curve to minimize the potential impact of LET. A multi-step compensator (as used in previous experiments[34, 37-39]) can be designed to study the synergistic effects of LET and FLASH dose rate in the future. For example, the distal edge of a proton Bragg curve is usually placed to normal tissue in the treatment, where the biological effect is increased by the high LET[34], inducing a higher risk of toxicity such as necrosis[42], however the FLASH effect might alleviate this issue[43]. Investigating the interplay of LET (increasing damage) and FLASH dose rate (sparing normal tissue) in biological effects is a worthwhile research endeavor. Second, the current study involved using a first scatterer to broaden the beam, with a small-hole collimator used to shape the beam profile to conform to the shape of the target. Use of these beam-shaping devices reduced the dose rate to ~100 to 200 Gy/s (**Figures 13** and **14**). To investigate dose rates beyond 1,000 Gy/s (**Figures 11** and **12**), we could use the original narrow beam without the first scatterer and any beam-shaping devices; in other words, a scan pattern could be optimized through the superposition of multiple original narrow beams to form a large irradiation field. The use of the small beamlet with higher dose rates (>1,000 Gy/s) has great potential for application in proton FLASH-based radiosurgery. Third, installation of a robotic motion and rotation platform equipped with high-precision imaging guidance (e.g., cone-beam CT and bioluminescence tomography[44]) would allow proton FLASH-based IMPT for small-animal irradiation experiments.

The significance of this study lies in its demonstration of our successful strategy to deliver proton FLASH to larger fields and volumes. The novelty is in applying an automatic robotic motion platform to accurately control the motion of the irradiation target (either 96-well plates or mice). The ultimate value of the present work is in paving the way for investigations of the radiobiological mechanisms underlying the effects of proton FLASH.

## 5. CONCLUSIONS

We validated the feasibility of accurately irradiating preclinical samples with FLASH proton spots by using a robotic motion platform and an external laser positioning system. The successful completion of this work has paved the way for preclinical studies to improve our understanding of the mechanisms underlying proton FLASH effects on normal tissue sparing. The methods developed here can be extended to other studies, including the synergistic effects of LET and dose rate, and the implementation of much higher dose rate (>1,000 Gy/s) proton FLASH-based IMPT. The superior physics dose distribution and potential radiobiological advantages of proton FLASH therapy enhance the inherent ability of radiotherapy to control tumors while providing further sparing of normal tissues, with the ultimate goal of improving clinical outcomes.

**Acknowledgments**

We thank Mr. Zhipeng Du from Rotrics for guidance in developing the G-code of the robotic motion platform. We thank Mr. Kelly Tharp and Mr. Paul Wisdom from the Department of Radiation Physics at MD Anderson Cancer Center for fabricating the accessories and parts for the experiments. We thank Ms. Christine F. Wogan from the Division of Radiation Oncology at MD Anderson Cancer Center for editorial contributions to this report. This work was supported by the National Cancer Institute [award number P30 CA016672] and by Radiation Oncology Strategic Initiatives Boot Walk Award from MD Anderson Cancer Center. The authors also acknowledge the Texas Advanced Computing Center at The University of Texas at Austin (http://www.tacc.utexas.edu) for providing high-performance computation resources that contributed to the research results reported within this paper. URL: http://www.tacc.utexas.edu

**Conflict of Interest**

Kiminori Iga is currently employed by Hitachi Ltd. Other authors declare no conflicts of interest.

**Institutional Review Board Statement**

Not applicable.

**Informed Consent Statement**

Not applicable.

## Data Availability Statement

The experimental data generated in the current study are available from the first author upon reasonable request.